\documentclass[aps,reprint,amsmath,amssymb,pre,floatfix]{revtex4-2}

\usepackage{multirow} 
\usepackage{xcolor}
\usepackage{graphicx}
\usepackage{dcolumn}
\usepackage{bm}
\usepackage{makecell}
\usepackage{placeins}
\usepackage{array}
\newcolumntype{P}[1]{>{\raggedright\arraybackslash}p{#1}}

\begin{document}
  
\title{Deep learning for classifying dynamical states from time series via recurrence plots}

\author{Athul Mohan${}^{1}$, G. Ambika${}^{2}$, Chandrakala Meena${}^{1}$}
\affiliation{${}^{1}$ Indian Institute of Science Education and Research (IISER), Pune, India}
\affiliation{${}^{2}$ Indian Institute of Science Education and Research (IISER), Thiruvananthapuram, India}

\date{\today} 

\begin{abstract}
Recurrence Quantification Analysis (RQA) is a widely used method for capturing the dynamical structure embedded in time series data, relying on the analysis of recurrence patterns in the reconstructed phase space via recurrence plots(RPs). Although RQA proves effective across a range of applications, it typically requires the computation of multiple quantitative measures, making it both computationally intensive and sensitive to parameter choices. In this study, we adopt an alternative approach that bypasses computation of recurrence measures, by directly using images of RP as input to a deep learning model. We propose a new dual-branch deep learning model named as DBResNet-50 build up on the ResNet-50 architecture and  specifically designed to efficiently capture the complex dynamical features encoded in RPs. We compare its performance with standard ResNet-50 and MobileNetV2 and find that DBResNet-50 consistently outperforms both benchmark models. Our DBResNet-50 model, trained exclusively on simulated time series, accurately classifies seven dynamical regimes—periodic, quasi-periodic, chaotic, hyperchaotic, white noise, pink noise, and red noise while consistently outperforming ResNet-50 and MobileNetV2. Further to assess its generalizability, we test the trained model to RP images generated from standard dynamical systems, which are not included in the training set, as well as to experimental datasets from a Chua circuit, X-ray light curves from the black-hole system GRS 1915+105, and observational light curves of the variable stars AC Her, SX Her, and Chi Cygni. In all cases, DBResNet-50 outperforms and correctly predicts the known dynamics of these systems. The model further infers the relative contributions of deterministic and stochastic components within a signal, as we observe through the analyses of temperature data from Ladakh and Ranchi locations. These results further demonstrate the robustness and versatility of our deep learning framework, underscoring the potential of RP image based models as fast, accurate, and scalable tools for classifying dynamical states in both synthetic and real-world time series data.

\end{abstract}

\maketitle

\maketitle

\section{\label{sec:introduction}Introduction}
Understanding and classifying emergent dynamical states in natural and engineered systems remains a fundamental challenge due to the intrinsic complexity of their physical mechanisms and the involvement of multiple interacting parameters. However, the increasing availability of large scale time series data and advances in high-performance computing open new opportunities to uncover meaningful dynamical patterns by integrating data driven approaches with computational modeling. Many real-world systems such as biological, climate, and social systems, and engineered systems, such as mechanical, electronics and power grid systems often generate nonlinear time series that encode a variety of rich dynamical behaviors, including periodic, quasiperiodic, chaotic, and hyperchaotic regimes \cite{Kantz_Schreiber_2003, meena2017effect,crutchfield1986chaos, meena2020resilience,lakshmanan1996chaos}.
Accurate identification of these behaviors is essential not only for advancing our understanding of the underlying physical processes but also for anticipating critical transitions, deriving early warning signals, and designing robust prediction and control strategies \cite{scheffer2009early, lenton2011early, meena2017threshold, scholl2008handbook}.

Traditionally, the classification of time series from dynamical systems relies on techniques such as Lyapunov exponent spectra \cite{bespalov2013determination,wolf1985determining}. However, this approach becomes impractical in the presence of stochasticity or when the governing equations do not fully capture the underlying dynamics of the system. Moreover, estimating Lyapunov exponents from empirical data requires large volumes of clean stationary signals, which limits their applicability in real-world scenarios \cite{lyapunovzou}. Thus, developing methods to capture the inherent dynamics of a system from empirical time series data remains a challenging and active area of research.

Recently Recurrence Quantification Analysis (RQA) has emerged as one of the most powerful techniques for uncovering dynamical states from time series data, whether derived from dynamical models or empirical measurements. RQA has been successfully applied across a wide range of disciplines, for example, in biological systems for the analysis of heart rate variability analysis \cite{PhysRevE.66.026702} and neural activity characterization \cite{ramdani2021parametric}; in climate science to identify climate change dynamics \cite{zhao2011identifying, trauth2019classifying,john2024recurrence}; and in engineering systems for the prediction of mechanical failures \cite{qian2012damage}.
This method analyzes recurrence patterns in the reconstructed phase space from data using recurrence plots (RPs).  Introduced by Eckmann et al. \cite{eckmann1995recurrence}, RPs provide a visual representation of the times when the state of the system returns close to the states visited previously\cite{Ambika2020}. They reveal complex structure related to dynamical features, such as periodicity, laminar phases, drifts, and chaotic bursts, that often escape detection by traditional linear or spectral methods \cite{marwan2007recurrence, kugiumtzis1998regularized}. The RQA extends the basic recurrence plot framework by introducing a set of quantitative measures, such as recurrence rate, determinism, laminarity, and entropy, that enable systematic analysis of the underlying dynamics \cite{webber2015rqa}.  However, as demonstrated in a recent study \cite{thakur2024mlrecurrence}, these measures often exhibit overlapping value ranges across different dynamical regimes, making it difficult to uniquely identify a specific state based on any single measure. Thus, reliable classification, typically requires a carefully chosen combination of multiple RQA features, underscoring the limitations of relying on individual measures for accurate identification of dynamical states.

With recent progress in machine learning, a new research direction has emerged that integrates the recurrence based measures with machine learning algorithms for automated classification of dynamical regimes. Several studies have used RQA derived features as input to standard machine learning or deep learning models that demonstrated improved classification performance \cite{thakur2024mlrecurrence, mathunjwa2021ecg}. However, the computational cost of extracting RQA features, especially for long or high dimensional time series, remains a significant limitation \cite{rawald2014fast}. To mitigate this, faster alternatives have been proposed, such as parametric RQA \cite{ramdani2021parametric} and symbolic recurrence quantification \cite{caballero2018symbolic}.

An alternative approach would be feeding the time series directly into a machine learning model, but they primarily encode statistical correlations and do not reveal the underlying state-space geometry, such as phase space structure or evolution of trajectory explicitly \cite{silveira2025classifying}. Moreover, although deep learning models, in principle, can learn internal state representations, they do so in a data-intensive and opaque manner and typically without physical interpretability. In addition, direct use of time series data leads to lower classification accuracy and substantially higher computational cost compared to recurrence based features, as demonstrated in \cite{silveira2025classifying}.
In contrast, the RPs yield meaningful, and interpretable measures that directly reflect the dynamical properties of the systems. Therefore, we rely on RP images directly to classify dynamical states more accurately and with clearer physical insights.

A promising novel approach that bypasses the need for explicit computation of RQA measures for feature extraction altogether would be feeding recurrence plots directly into deep learning models, particularly convolutional neural networks (CNNs). The CNNs, widely used in image classification, automatically learn hierarchical spatial features through convolutional layers. This image based strategy enables CNNs to extract discriminative patterns directly from recurrence plots without requiring prior feature engineering. A few notable applications in this context include arrhythmia detection from ECG data \cite{mathunjwa2021ecg}, activity recognition using wearable sensors \cite{garciaceja2018activity}, control system development \cite{fradkov1998control}, and parameter estimation in nonlinear systems \cite{piva2024predictive}. These studies show that RP based deep learning models can match or even outperform traditional recurrence measure-based approaches, offering greater scalability and adaptability. Despite these advances, CNN architectures suffer from vanishing gradients, which hinder the training of deeper models. This challenge is largely resolved with the introduction of ResNet architectures \cite{resnet50}, which use residual connections to preserve gradient flow and enable deep network training.  

In our study, we introduce a dual-branch deep learning model named DBResNet-50, which incorporates ResNet-50\cite{resnet50} as the backbone for feature extraction from images of RPs, followed by two parallel branches inspired by the DSRNet architecture \cite{zhai2020dsrnet}. These branches are designed to emphasize features corresponding to various directional edges, which are highly relevant for RPs. We then evaluate the performance of our proposed model along with the standard ResNet-50. We also consider the MobileNetV2 model \cite{sandler2018mobilenetv2}, which is a lighter architecture with fewer parameters than ResNet-50.

All three models are trained on recurrence plots to classify time series into seven categories—periodic, quasi-periodic, chaotic, hyperchaotic, white noise, pink noise, and red noise, with the noise types chosen due to their prevalence in natural systems. Then, we compare their performances on a variety of data sets, derived from synthetic and real world systems. The results show that our dual-branch architecture achieves the highest accuracy while maintaining superior computational efficiency.

Our paper is organized as follows. In section~\ref{sec:methodology}, we discuss the methodology for the classification of nonlinear time series by integrating machine learning models and recurrence plots.  Section~\ref{sec:results} presents the results that show the performance of the deep learning models for simulation and real-world data sets. Section~\ref{sec:Discussion} discusses the applicability as well as the limitations of our approach. In Section~\ref{sec:conclusion}, we summarize the results and indicate promising future extensions for our model. 

\section{\label{sec:methodology}Methodology}
To generate the training and testing datasets, we first simulate a set of standard dynamical systems along with 3 types of noise, as summarized in Table~\ref{tab:dynamical_systems}. From the resulting time series, we reconstruct the trajectories in phase space using Takens' embedding theorem and derive the corresponding RPs to capture the intrinsic recurrence structure of the underlying dynamics.

\begin{figure*}[!ht]
    \centering
    \includegraphics[width=0.75\linewidth]{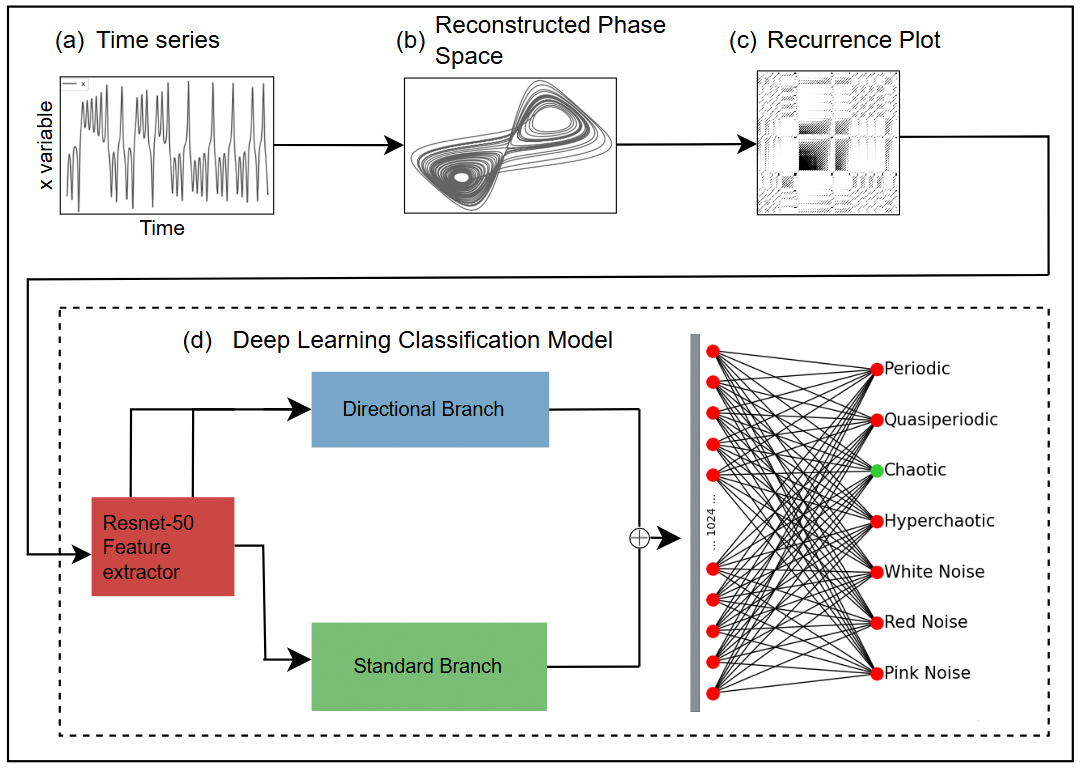}
    \caption{\textbf{Classification of time series.} Schematic representation of the proposed methodology for classifying time series using RPs as input to the proposed deep learning model. (a) We begin by selecting a normalized time series from some simulated dynamical systems and noise generator as listed in Table~\ref{tab:dynamical_systems}. (b) The phase space of the selected time series is reconstructed using Takens’ embedding theorem. (c) A recurrence plot is then generated from the embedded trajectory using a suitable recurrence threshold $\epsilon$. (d) The resulting RP is resized and used as the input to the deep learning model. In this model, ResNet-50 helps to extract features from RP. Its output is processed through two parallel branches. The standard branch captures high-level spatial features, and the directional branch uses multiple early feature maps from the ResNet 50 pipeline, which are shown by the two arrows drawn from the mid-layers of ResNet 50. The outputs from both branches are concatenated and passed through a dense layer. Finally, a softmax layer generates class wise probabilities, from which the most probable dynamical state is identified. For example, if the input time series is chaotic, the output layer assigns a higher probability to the chaotic class.
}
    \label{fig:pipeline}
\end{figure*}

These RPs are then used as input features for the proposed deep learning framework, which is designed to classify distinct dynamical regimes and noise types. A schematic representation of the basic methodology that indicates the systematic way of classification of time series using the dual branch deep learning model is shown in Fig.~\ref{fig:pipeline}. Comprehensive details on time series generation, RP construction, and the architecture of the proposed deep learning model are provided in the following subsections.

\subsection{\label{sec:timeseries}Time series datasets}

To generate the time series datasets, we simulate standard nonlinear dynamical systems as listed in Table~\ref{tab:dynamical_systems} using the RK45 integrator. By systematically varying the model parameters and initial conditions, we ensure a broad coverage of dynamical behaviors. For instance, Chen 4D and Lorenz 4D systems are used to generate time series exhibiting chaotic, hyperchaotic, quasi-periodic, and periodic dynamics for choosing appropriate parameters. The Lorenz 3D system is used to generate an additional chaotic dataset. We use the generator circuit dynamics in Table~\ref{tab:dynamical_systems} to generate time series that show quasi-periodic behavior. For each simulated system, we use a single time series corresponding to the $x$ variable (see Fig. \ref{fig:pipeline}(a)). In addition, we generate noisy signals of Gaussian white noise,  pink noise, and red noise, using the algorithm described by Timmer and Koenig\cite{timmerandkoenig}. In the deterministic classes, we produce noisy datasets by adding Gaussian noise to the state variable $x$. The standard deviation of the noise is set as a fixed proportion of the original data, as described below:
\begin{equation}
\begin{aligned}
x_{\text{noisy}}(t) &= x(t) + \mathcal{N}\left(0, \sigma_n^2\right), \\
\text{with} \quad \sigma_n &= \alpha \cdot \sigma_x, \quad \alpha \sim \mathcal{U}(0, 0.05), \\
\end{aligned}
\label{eq:noise}
\end{equation}

where \( \sigma_x \) denotes the standard deviation of the data \( x(t) \) obtained from the dynamical system, and \( \sigma_n \) denotes the standard deviation of noise.
We normalize each dataset before using it to reconstruct the phase space trajectory and analyze recurrence features, as detailed in the following section.
\begin{table*}[t]
\centering
\caption{Details of dynamical and stochastic systems used to generate different classes of time series data for training and testing.}
\label{tab:dynamical_systems}

\begingroup
\scriptsize
\setlength{\tabcolsep}{4pt}
\renewcommand{\arraystretch}{1.05}

\begin{tabular}{
|>{\centering\arraybackslash}p{2.5cm}
|>{\centering\arraybackslash}p{4.5cm}
|>{\centering\arraybackslash}p{2.3cm}
|>{\centering\arraybackslash}p{1.8cm}|
}
\hline
\textbf{System}
&
\textbf{Equations / Algorithms}
&
\textbf{Parameters}
&
\textbf{Class}
\tabularnewline
\hline\hline

Chen 4D\cite{chen2007hyperchaos}
&
\begin{tabular}[t]{@{}l@{}}
$\dot{x}=a(y-x)+e\,yz$ \\[3pt]
$\dot{y}=cx-d\,xz+y+u$ \\[3pt]
$\dot{z}=xy-bz$ \\[3pt]
$\dot{u}=-k\,y$
\end{tabular}
&
\begin{tabular}[t]{@{}l@{}}
$a=35,\;c=25$ \\[2pt]
$d=5,\;e=35$ \\[2pt]
$k=100,\;b=7.6$ \\[2pt]
$k=100,\;b=4.9$ \\[2pt]
$k=315,\;b=4.9$ \\[2pt]
$k=195,\;b=4.9$ \\[2pt]
$k=495,\;b=4.9$ \\[2pt]
$k=800,\;b=4.9$ \\[2pt]
$k=590,\;b=4.9$
\end{tabular}
&
\begin{tabular}[t]{@{}l@{}}
\mbox{} \\[2pt]
\mbox{} \\[2pt]
Hyperchaotic \\[2pt]
Hyperchaotic \\[2pt]
Chaotic \\[2pt]
Chaotic \\[2pt]
Periodic \\[2pt]
Quasiperiodic \\[2pt]
Quasiperiodic
\end{tabular}
\tabularnewline
\hline

Lorenz 4D\cite{li2011hyperchaotic}
&
\begin{tabular}[t]{@{}l@{}}
$\dot{x}=a(y-x)$ \\[3pt]
$\dot{y}=bx-xz-cy+w$ \\[3pt]
$\dot{z}=xy-dz$ \\[3pt]
$\dot{w}=-k\,y-r\,w$
\end{tabular}
&
\begin{tabular}[t]{@{}l@{}}
$a=12,\;b=23$ \\[2pt]
$c=1,\;d=2.1$ \\[2pt]
$k=4.00,\;r=0.20$ \\[2pt]
$k=6.00,\;r=0.10$ \\[2pt]
$k=5.00,\;r=0.20$ \\[2pt]
$k=6.00,\;r=1.00$ \\[2pt]
$k=0.10,\;r=0.20$ \\[2pt]
$k=0.30,\;r=0.20$ \\[2pt]
$k=6.00,\;r=0.45$ \\[2pt]
$k=20.0,\;r=0.20$
\end{tabular}
&
\begin{tabular}[t]{@{}l@{}}
\mbox{} \\[2pt]
\mbox{} \\[2pt]
Hyperchaotic \\[2pt]
Hyperchaotic \\[2pt]
Hyperchaotic \\[2pt]
Chaotic \\[2pt]
Chaotic \\[2pt]
Chaotic \\[2pt]
Periodic \\[2pt]
Periodic
\end{tabular}
\tabularnewline
\hline

Lorenz 3D\cite{Ghys2013}
&
\begin{tabular}[t]{@{}l@{}}
$\dot{x}=\sigma(y-x)$ \\[3pt]
$\dot{y}=x(\rho-z)-y$ \\[3pt]
$\dot{z}=xy-\beta z$
\end{tabular}
&
\begin{tabular}[t]{@{}l@{}}
$\sigma=10,\;\beta=8/3$ \\[3pt]
$\rho=28.00$
\end{tabular}
&
\begin{tabular}[t]{@{}l@{}}
\mbox{} \\[3pt]
Chaotic
\end{tabular}
\tabularnewline
\hline

Generator circuit dynamics\cite{kuznetsov2023coupled}
&
\begin{tabular}[t]{@{}l@{}}
$\dot{x}=y$ \\[3pt]
$\dot{y}=(\lambda+z+x^2-\beta x^4)y-\omega_0^2x$ \\[3pt]
$\dot{z}=b(c-z)-ky^2$
\end{tabular}
&
\begin{tabular}[t]{@{}l@{}}
$\lambda=-1,\;c=4$ \\[2pt]
$\beta=\tfrac{1}{18},\;\omega_0=2\pi$ \\[2pt]
$b=1$ \\[2pt]
$k=0.02$
\end{tabular}
&
\begin{tabular}[t]{@{}l@{}}
\mbox{} \\[4pt]
Quasiperiodic
\end{tabular}
\tabularnewline
\hline

Stochastic Data\cite{timmerandkoenig}
&
\begin{tabular}[t]{@{}l@{}}
$N_1,N_2\sim\mathcal{N}(0,1)$ \\[3pt]
$S(f_k)\sim f_k^{-\beta}$ \\[3pt]
$F(f_k)=\sqrt{\dfrac{S(f_k)}{2}}\,(N_1+iN_2)$ \\[3pt]
$x(t)=\mathcal{F}^{-1}\!\left[F(f_k)\right]$
\end{tabular}
&
\begin{tabular}[t]{@{}l@{}}
$\beta=0$ \\[3pt]
$\beta=1$ \\[3pt]
$\beta=2$
\end{tabular}
&
\begin{tabular}[t]{@{}l@{}}
White Noise \\[3pt]
Pink Noise \\[3pt]
Red Noise
\end{tabular}
\tabularnewline
\hline

\end{tabular}

\endgroup
\end{table*}

\subsection{\label{sec:reconstruction}Reconstruction of Phase space}
From the normalized univariate time series, we reconstruct the system’s phase space using Takens’ embedding theorem~\cite{doi:10.1142/S0218127491000634}. According to this theorem, the dynamical structure of the original multi-dimensional state space can be recovered by constructing delay vectors with appropriate delay and embedding dimension as, 
\begin{equation}
\begin{aligned}
\mathbf{x}(t) = (&x(t), x(t+\tau), x(t+2\tau), \dots, \\
                &x(t+(d-1)\tau))
\end{aligned}
\label{eq:embedding}
\end{equation}

where \( \tau \) is the embedding delay, and \( d \) is the embedding dimension.

Various techniques are available to estimate appropriate values for the time delay $\tau$ and the embedding dimension $d$, which are essential to reconstruct the trajectory of the system from a single scalar time series. In our study, we employ the mutual information method \cite{tan2023embedding} to estimate $\tau$. The first local minimum of the mutual information curve is typically selected as the optimal delay, as it marks the point at which successive values of the time series become sufficiently independent for effective phase space reconstruction. Although the precise value of the delay may depend on the resolution of the time series, the method remains robust provided that the search is performed over a sufficiently large range to capture the local minimum.
To determine the embedding dimension $d$, we use the method of False Nearest Neighbors (FNN), following the approach proposed by Kennel et al. \cite{kennel1992embedding}. The underlying idea is that if the reconstructed trajectory is folded onto itself in a lower-dimensional space, it results in a higher number of false neighbors. The knee point at which this number levels off to zero indicates a suitable embedding dimension that preserves the system's topology.

With the appropriate choice of \( \tau \) and \( d \) for each time series sample from the systems, the original univariate time series is transformed into a sequence of embedded vectors. These vectors are organized into a matrix \( \mathbf{X} \) as follows:

\begin{equation}
\mathbf{X} =
\begin{bmatrix}
x_1 & x_{1+\tau} & \cdots & x_{1+(d-1)\tau} \\
x_2 & x_{2+\tau} & \cdots & x_{2+(d-1)\tau} \\
x_3 & x_{3+\tau} & \cdots & x_{3+(d-1)\tau} \\
\vdots & \vdots & \ddots & \vdots \\
x_N & x_{N+\tau} & \cdots & x_{N+(d-1)\tau}
\end{bmatrix}
\label{eq:embedding_matrix}
\end{equation}
 where each row represents a delay vector or an embedded vector, and plotting these vectors in the reconstructed phase space reveals the underlying structure of the system, capturing its essential dynamical properties. Thus, this matrix effectively reconstructs the phase space trajectory and serves as the basis for subsequent recurrence analysis.

\subsection{\label{sec:recurrence plots}Recurrence Plots}
Recurrence plots provide a two-dimensional representation of the times when a dynamical system revisits states that are close to those encountered previously, within a predefined distance threshold $\epsilon$. Given the delay embedded matrix \( \mathbf{X} \) in Eq. \ref{eq:embedding_matrix}, the recurrence matrix \( \mathbf{R} \) is constructed with elements:
\begin{equation}
\begin{split}
R_{i,j} &= \Theta(\varepsilon - \|\mathbf{x}_i - \mathbf{x}_j\|), \\
\text{where } \Theta(u) &=
\begin{cases}
1 & \text{if } u \geq 0 \\
0 & \text{if } u < 0
\end{cases}
\end{split}
\end{equation}

The recurrence plot is then derived by adding dots corresponding to the nonzero elements of R. Choosing an appropriate threshold $\epsilon$ is critical for constructing meaningful recurrence plots, as it determines which state pairs are considered recurrent and thereby influences the actual dynamical state of the system. An optimal threshold preserves significant features of the dynamics while excluding noise-induced or irrelevant recurrences \cite{marwan2011pitfalls}. 
If $\epsilon$ is too small, the recurrence plot may contain very few recurrence points, leading to the omission of important dynamical patterns. Conversely, if $\epsilon$ is too large, the plot becomes overly dense, which can obscure important structures such as diagonal lines that reflect periodic or quasi-periodic behavior.

\begin{figure}
    \centering
    \includegraphics[width=1\linewidth]{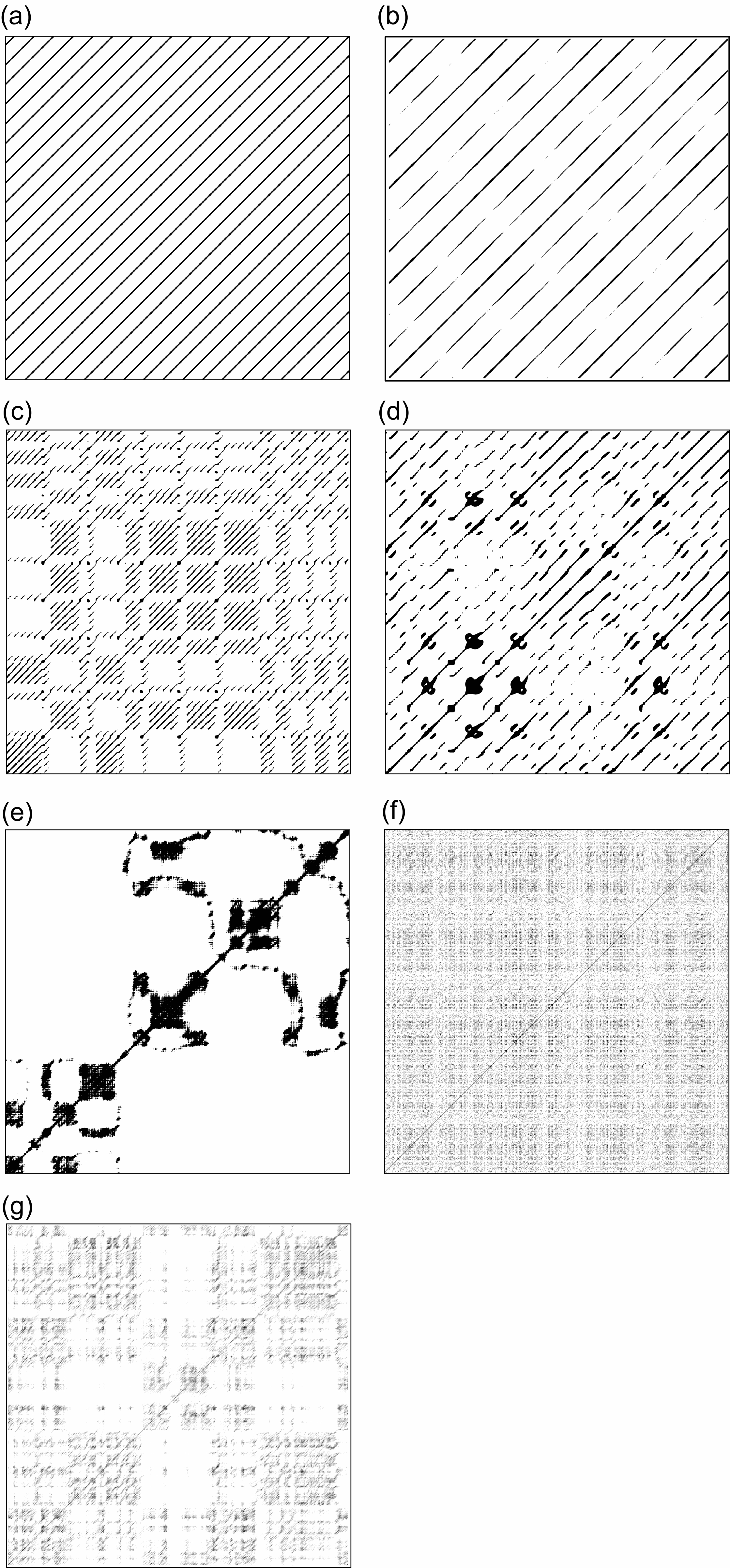}
    \caption{\textbf{Representative recurrence plots from all the seven classes}. These plots are obtained from the univariate time series corresponding to the following seven classes: (a) Periodic (Lorenz 4D), (b) Quasiperiodic, (c) Chaotic (Lorenz 4D), (d) Hyperchaotic (Chen 4D), (e) Red Noise, (f) White Noise, and (g) Pink Noise.}
    \label{RP_plots}
\end{figure}

To select an appropriate threshold, the common approaches include fixing $\epsilon$ as a percentage of the signal’s standard deviation \cite{ambrozkiewicz2019diagnostics}, or adjusting it to maintain a fixed recurrence rate across datasets \cite{marwan2007recurrence}.
In our study, we set the threshold $\epsilon$ to achieve a fixed recurrence rate of $5\%$. To do this, we compute all pairwise distances in the embedded d-dimensional phase space and select the distance value at which $5\%$ of the distances are smaller than or equal to it, as the threshold $\epsilon$. Fig.~\ref{RP_plots} presents representative RPs for each class generated from different dynamical systems and noise profiles.

\subsection{\label{sec:ml}Deep Learning Models for Classification}

We first implement a  deep learning model based on the ResNet-50 architecture. The model uses a ResNet-50~\cite{resnet50} architecture pre-trained on the ImageNet dataset~\cite{russakovsky2015imagenet} and is subsequently fine-tuned on RP images. The final feature maps from the backbone are passed through a Global Average Pooling (GAP) layer. Finally, a softmax output layer produces the class probability distribution. The model uses categorical cross-entropy loss and the AdamW\cite{llugsi2021comparison} optimizer, and training is performed with early stopping based on validation accuracy to prevent overfitting.

We also fine-tune a MobileNetV2 architecture pre-trained on ImageNet and subsequently fine-tuned on the RP dataset, using the same transfer-learning setup as ResNet-50.
MobileNetV2 exhibits lower performance than ResNet-50, and the classification results produced by both models are not reliable across all data classes.

In this context, we introduce a dual-branch deep learning implementation based on ResNet-50, termed the DBResNet-50 model. The ImageNet-pretrained ResNet-50 backbone is used without its classifier. The standard branch applies global average pooling and $L2$ normalization to the final ResNet-50 features. The directional branch extracts features from the third convolutional layer in stage 2 and an upsampled fourth convolutional layer in stage 3, concatenates them, and processes them with a $1\times1$ convolution (128 channels, ReLU), a squeeze $1\times1$ convolution, eight fixed directional $3\times3$ kernels, batch normalization with ReLU, and three dilated $3\times3$ convolutions (dilations 1, 2, 3). Its output is globally averaged, batch-normalized, $L2$-normalized, and fed to an auxiliary softmax head. A learnable channel-wise gate scales this branch’s output before concatenation with the standard branch. The fused features pass through a 1024-unit fully connected layer (BN, ReLU, dropout = 0.5) and a final softmax layer.

Training minimizes $L = L_{\text{main}} + 0.3L_{\text{aux}}$ using AdamW (learning rate $5\times10^{-4}$, weight decay $10^{-4}$). Phase~1 trains only newly added layers for 12 epochs; Phase~2 fine tunes the backbone stages from \texttt{conv3} to \texttt{conv5} for up to 100 epochs, with early stopping set to stop training if the performance metric does not improve for 10 consecutive epochs. All models are trained with a batch size of 8.

\begin{figure}
    \centering
    \includegraphics[width=\linewidth]{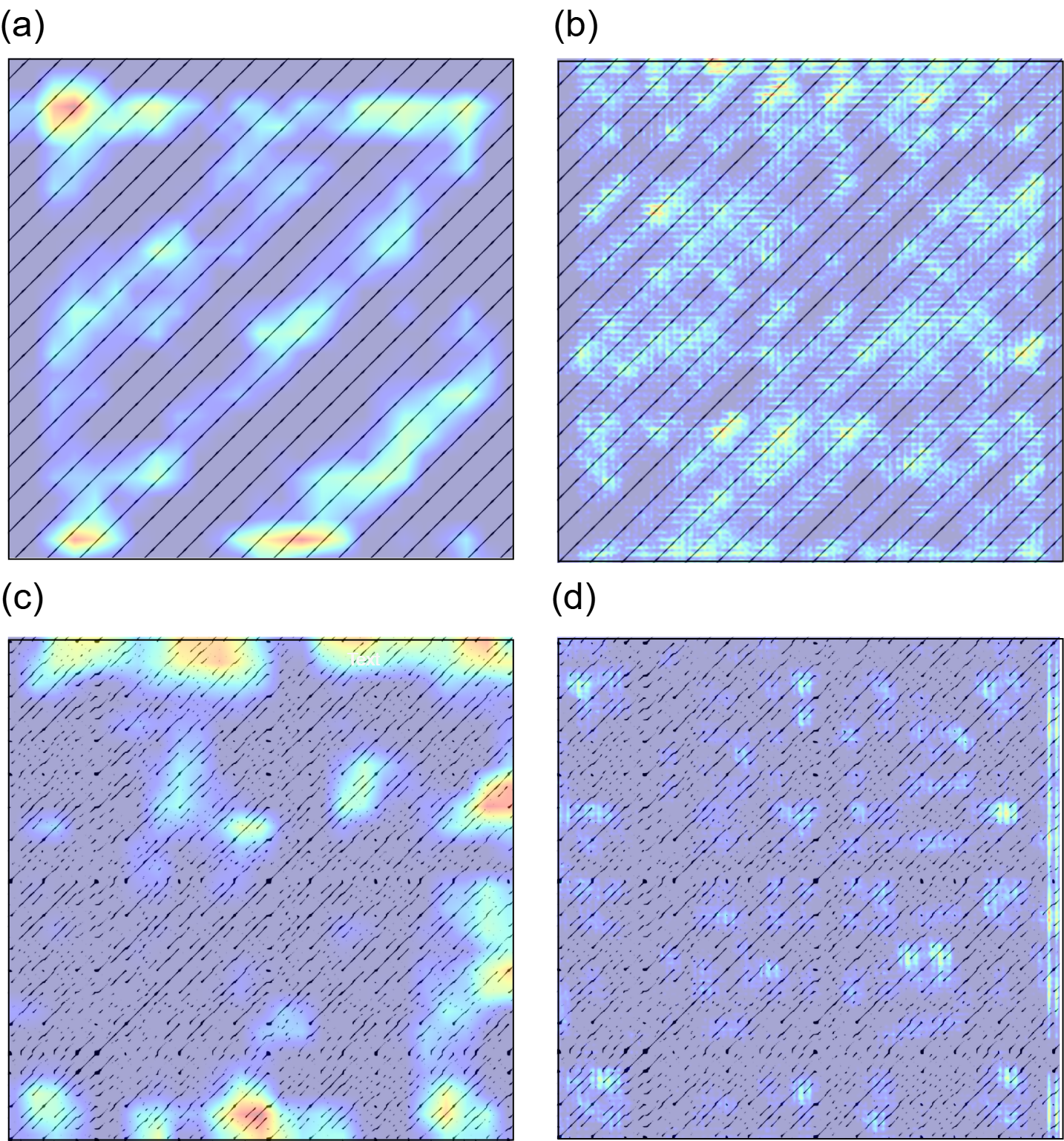}
    \caption{Grad-CAM visualizations of DBResNet-50 for the periodic and chaotic classes using representative RPs.} Panels (a) and (c) show the Grad-CAM heatmaps for the standard branch, whereas panels (b) and (d) show the corresponding heatmaps from the directional branch. Warmer colors indicate regions that contribute more strongly to the predicted class.
    \label{fig:gradcam}
\end{figure}

Furthermore, to interpret the features learned by the proposed model, we apply Grad-CAM \cite{mohamed2024enhancing} visualization to the standard branch and the directional branch. As shown in Fig.~\ref{fig:gradcam}, the standard branch predominantly attends to localized structures in the input RP, whereas the fused model occasionally highlights a broader set of discriminative regions. This suggests that the directional branch provides complementary information that enhances the representation learned by the standard branch.

\section{\label{sec:results}Results}

The main objective of this study is to accurately identify and classify the distinct dynamical behaviors and noisy patterns of time series, using a deep-learning model trained on RPs generated from reconstructed phase-space trajectories. To determine the importance of embedded RPs, we compare the performance of the proposed deep-learning framework using RPs generated with and without embedding. Under identical training and testing protocols, the embedded representation achieves an accuracy of $92.6\%$, compared with $71.5\%$ for the non-embedded representation. These results demonstrate that embedded RPs constructed provide additional discriminative information for classifying dynamical regimes within the proposed framework. Accordingly, throughout this work, all reported results are based on RP representations generated from embedded phase-space trajectories.

For training, we generate 600 RPs per class from various dynamical systems and noise models, as listed in Table \ref{tab:dynamical_systems}. For validation, we generate $100$ RPs per class from the systems in Table \ref{tab:dynamical_systems}, ensuring that they are distinct from the training set.

The complete data set contains seven distinct classes. A few typical representative RPs from each class are presented in Fig.~\ref{RP_plots}, with the images resized to $800 \times 800$ pixels.

 To enhance prediction accuracy, we apply a test-time augmentation strategy in which each input image is center-cropped by up to $15\%$, while preserving the original aspect ratio. Predictions are generated for all augmented variants of the same image, and the class with the highest average confidence across these predictions is selected as the final output. This ensemble-like approach improves robustness to minor variations in RP appearance.

To evaluate the performance of the classification models, we compute the confusion matrices and standard classification metrics \cite{Ting2010,rainio2024evaluation}. In addition, we determine the per-class accuracy, defined as follows:
Given a confusion matrix \( \mathbf{C} \in \mathbb{N}^{K \times K} \), where \( C_{i,j} \) denotes the number of samples whose true class is \( i \) and predicted class is \( j \), and \( K \) denotes the number of classes, the accuracy per class for class \( i \) is defined as:

\begin{equation}
\text{Accuracy}_i = \frac{C_{i,i}}{\sum_{j=1}^{K} C_{i,j}}, \quad \text{for } i = 1, 2, \dots, K
\end{equation}

\noindent
where, \( C_{i,i} \) is the number of correctly predicted samples for the class \( i \),
    \( \sum_{j=1}^{K} C_{i,j} \) is the total number of  samples belonging to the class \( i \).
This metric reflects how well the model performs for each class individually.

\subsection{\label{sec:simdata}Classification of simulated time series data}

To establish the performance of the trained machine learning models, namely, ResNet-50, MobileNetV2, and the proposed DBResNet-50, the models are tested on RPs generated from simulated time series data produced with a different parameter regime that preserves the same class dynamical state as the training dataset. The corresponding performance results for all three models are summarized in Table~\ref{tab:performance_comparison}. DBResNet-50 model achieves the highest accuracy of 96.86\%, while ResNet-50 achieves 90.14\% and MobileNetV2 achieves 86.86\%.

\begin{table*}[ht]
\centering
\caption{Performance comparison of models for classifying test data from images of their RPs}
\label{tab:performance_comparison}
\resizebox{\textwidth}{!}{%
\begin{tabular}{|l|cccc|cccc|cccc|}
\hline
\textbf{Class} & \multicolumn{4}{c|}{\textbf{DBResNet-50}} & \multicolumn{4}{c|}{\textbf{ResNet-50}} & \multicolumn{4}{c|}{\textbf{MobileNetV2}} \\
 & Precision & Recall & F1-Score & Accuracy & Precision & Recall & F1-Score & Accuracy & Precision & Recall & F1-Score & Accuracy \\
\hline
Chaotic & 0.9333 & 0.8400 & 0.8842 & 0.8400 & 0.8053 & 0.9100 & 0.8545 & 0.9100 & 0.8571 & 0.5400 & 0.6626 & 0.5400 \\
Hyperchaotic & 0.8621 & 1.0000 & 0.9259 & 1.0000 & 0.9490 & 0.9300 & 0.9394 & 0.9300 & 0.5839 & 0.9400 & 0.7203 & 0.9400 \\
Periodic & 1.0000 & 0.9800 & 0.9899 & 0.9800 & 0.7143 & 0.8500 & 0.7763 & 0.8500 & 0.9796 & 0.9600 & 0.9697 & 0.9600 \\
Pink Noise & 1.0000 & 1.0000 & 1.0000 & 1.0000 & 0.9901 & 1.0000 & 0.9950 & 1.0000 & 0.9252 & 0.9900 & 0.9565 & 0.9900 \\
Quasi-periodic & 1.0000 & 0.9600 & 0.9796 & 0.9600 & 0.9054 & 0.6700 & 0.7701 & 0.6700 & 1.0000 & 0.6600 & 0.7952 & 0.6600 \\
Red Noise & 1.0000 & 1.0000 & 1.0000 & 1.0000 & 1.0000 & 0.9500 & 0.9744 & 0.9500 & 0.9709 & 1.0000 & 0.9852 & 1.0000 \\
White Noise & 1.0000 & 1.0000 & 1.0000 & 1.0000 & 1.0000 & 1.0000 & 1.0000 & 1.0000 & 0.9706 & 0.9900 & 0.9802 & 0.9900 \\
\hline
\textbf{Accuracy} & \multicolumn{4}{c|}{0.9686} & \multicolumn{4}{c|}{0.9014} & \multicolumn{4}{c|}{0.8686} \\
\hline
\end{tabular}%
}
\end{table*}

\begin{figure*}
    \centering
    \includegraphics[width=1\linewidth]{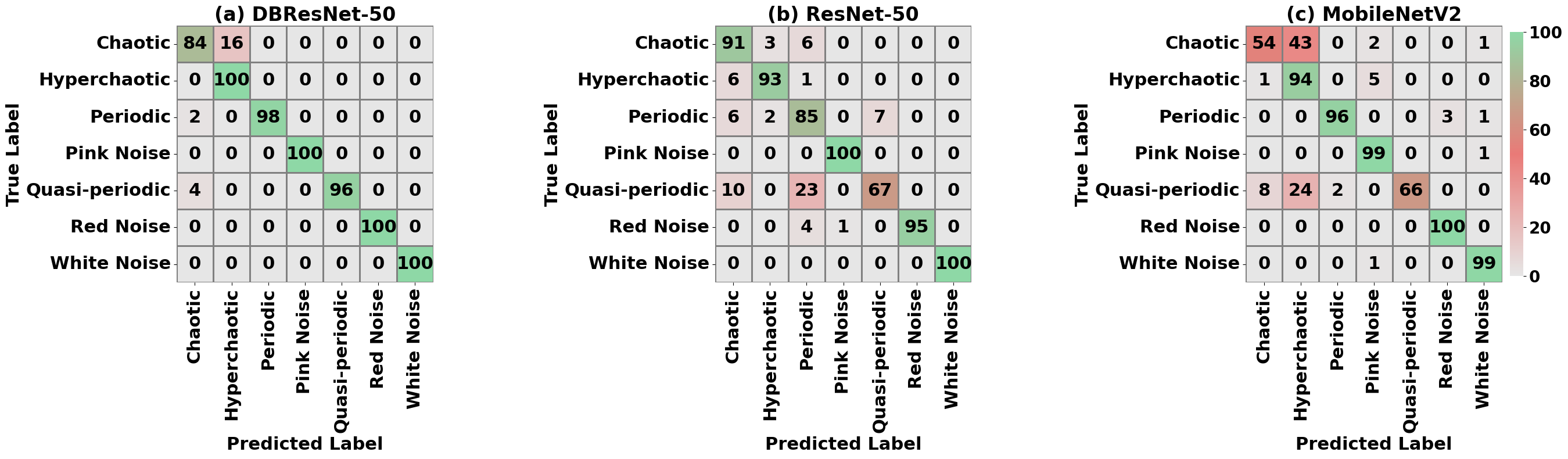}
    \caption{\textbf{Confusion matrices.} Confusion matrices showing classification performance of (a) DBResNet-50 (b) ResNet-50, (c) MobileNetV2 models on datasets generated by simulating models of Table~\ref{tab:dynamical_systems} at parameter sets different from training but show same type dynamical behaviour. Diagonal cells indicate correctly classified RPs, while off-diagonal cells indicate misclassifications. It is evident that DBResNet-50 is the best-performing model, as nearly all samples lie along the diagonal.}
    \label{fig:confusion matrix}
\end{figure*}

The confusion matrices in Fig.~\ref{fig:confusion matrix} present an overview of the classification performance across all seven classes for the three models. The diagonal elements indicate correct classifications, while off-diagonal elements represent misclassifications. It is evident that DBResNet-50 is the best-performing model, as nearly all samples lie along the diagonal.
The inference-time assessment indicates no considerable difference among the three models, with single-image inference times of 26.84 ms for MobileNetV2, 28.48 ms for ResNet-50, and 28.85 ms for DBResNet-50.

\begin{table*}
\centering
\caption{Details of additional dynamical systems used for generating data for validation.}
\label{tab:dynamical_systems_extended}
\scriptsize
\renewcommand{\arraystretch}{1.25}
\setlength{\tabcolsep}{4pt}

\begin{tabular}{|p{3.8cm}|p{4.2cm}|p{4.2cm}|p{1.8cm}|}
\hline
\textbf{System} & \textbf{Equations} & \textbf{Parameters} & \textbf{Class} \\
\hline\hline

R\"ossler System~\cite{ibrahim2018rossler} &
\(\begin{aligned}[t]
\dot{x} &= -y - z \\
\dot{y} &= x + ay \\
\dot{z} &= b + z(x-c)
\end{aligned}\) &
\(\,a=0.2,\; b=0.2,\; c=6\,\) &
Chaotic \\
\hline

Duffing System~\cite{korsch2008duffing1} &
\(\begin{aligned}[t]
\dot{x} &= y \\
\dot{y} &= x - x^{3} - \delta y + A\sin(\omega t)
\end{aligned}\) &
\(\,\delta=0.5,\; \omega=1,\; A=0.7\,\) &
Chaotic \\
\hline

Hyperchaotic System~\cite{benkouider2024comprehensive} &
\(\begin{aligned}[t]
\dot{x} &= ax - yz + w \\
\dot{y} &= xz - by \\
\dot{z} &= xy - cz \\
\dot{w} &= -y + d
\end{aligned}\) &
\(\,a=8,\; b=40,\; c=10,\; d=-0.1\,\) &
Hyperchaotic \\
\hline

\end{tabular}
\end{table*}

To further assess the robustness and generalization capability of the DBResNet-50 model, we evaluate its performance on RPs generated from dynamical systems not encountered during training. The governing equations and parameter values of these validation systems are summarized in Table~\ref{tab:dynamical_systems_extended}. For each system, we generate ten time series by varying the initial conditions while keeping the system parameters fixed. The model correctly predicts the underlying dynamics for all test datasets.

We additionally assess the models' robustness using surrogate data analysis \cite{neves2017recurrence}. For this purpose, we generate shuffled surrogates for all deterministic classes: periodic, quasiperiodic, chaotic, and hyperchaotic  using the test datasets listed in Table \ref{tab:dynamical_systems}. All three deep-learning models predominantly classify the surrogate time series as stochastic, demonstrating their sensitivity to the disruption of the underlying dynamical structure introduced by shuffled-surrogate randomization. Among the three architectures, ResNet-50 exhibits the strongest accuracy, assigning 99.7\% probability to the stochastic class. DBResNet-50 follows closely with 99\%, while MobileNetV2 assigns 97.4\% probability. These results indicate that the ResNet-50 backbone provides particularly robust classification for randomized surrogate signals, with DBResNet-50 retaining nearly the same level of robustness while incorporating its directional-branch architecture.

\subsection{\label{sec:realworlddata}Classification of empirical/observational time series data}

To further evaluate the generalizability of deep learning models, we test them on experimental or observational time series data. We use interpolation and binning strategies to address missing data and the irregular sampling of real-world time series.

\begin{figure}[ht]
    \centering
    \includegraphics[width=\linewidth]{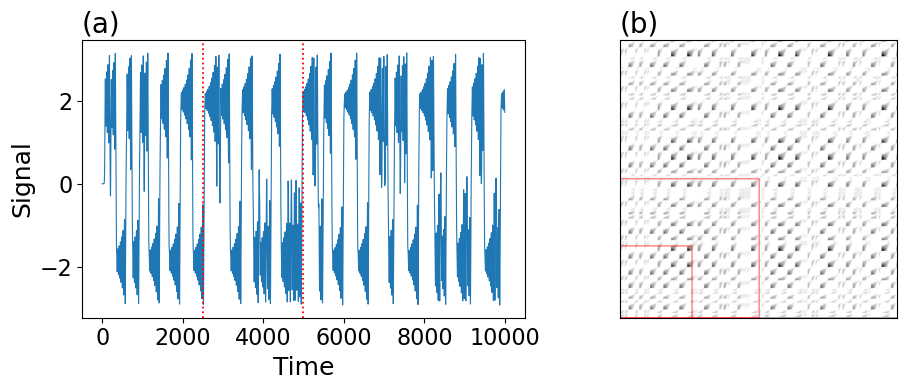}
    \caption{\textbf{Experimental data from Chua’s circuit.} 
    (a) Time series obtained experimentally from Chua’s circuit. The red lines indicate the cropped portions of the time series at 2,500 and 5,000 data points, while the entire plot corresponds to 10,000 points. 
    (b) Corresponding RP, with the red lines highlighting the regions in the RP corresponding to the cropped portions shown in (a).}
    \label{fig:chuas}
\end{figure}

As a first example, we use time series data obtained experimentally from the Chua circuit. We generate and pass the RPs with three different lengths, $10000, 5000$, and $2500$, of time series to the models to test their performance on varying pattern scales within the RPs. The cropping and the corresponding RPs are shown in Fig.~\ref{fig:chuas}. The DBResNet-50 model correctly classifies the Chua circuit signal as chaotic for all three cropped ensembles, consistent with the well-established chaotic dynamics of the Chua system~\cite{lakshmanan1996chaos, kuznetsov2023hidden}. The predictions of the other two models are not correct in all cases as is clear from Table \ref{tab:rr5_model_comparison_accuracy}.

\begin{figure}[ht]
    \centering
    \includegraphics[width=\linewidth]{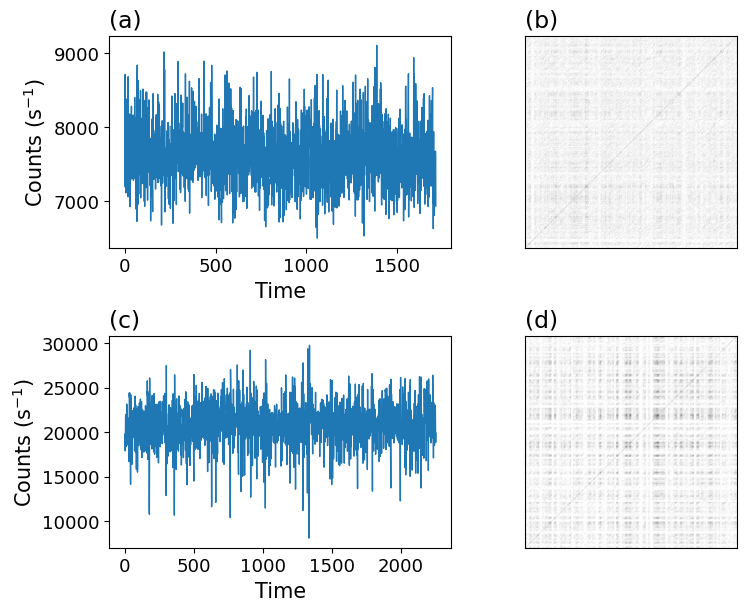}
    \caption{\textbf{X-ray light curves from the black hole GRS 1915+105.} 
    (a) Light curve data from category $\gamma$ and (b) its corresponding RP. 
    (c) Light curve data from category $\chi$ and (d) its corresponding RP.}
    \label{fig:GRS}
\end{figure}

Next, we validate the models using X-ray light-curve data from the black-hole system GRS~1915+105, obtained from the RXTE TOO archive (\url{https://heasarc.gsfc.nasa.gov}). The source is reported to exhibit twelve well-known variability classes, each corresponding to a distinct dynamical state characterized by its count-rate evolution and spectral color properties \cite{belloni2000model}. The data are processed as described in \cite{harikrishnan2011nonlinear}.
Among these variability classes, the $\gamma$ and $\chi$ classes are known to exhibit white noise behavior \cite{belloni2000model,harikrishnan2011nonlinear}. We select representative data of the light curves from these two classes, and the corresponding light curves and RPs are shown in Fig.~\ref{fig:GRS}. 
The $\chi$ class is correctly classified by all three models, whereas the $\gamma$ class is correctly predicted only by our DBResNet-50 model, as summarized in Table~\ref{tab:rr5_model_comparison_accuracy}.

\begin{figure}[ht]
    \centering
    \includegraphics[width=\linewidth]{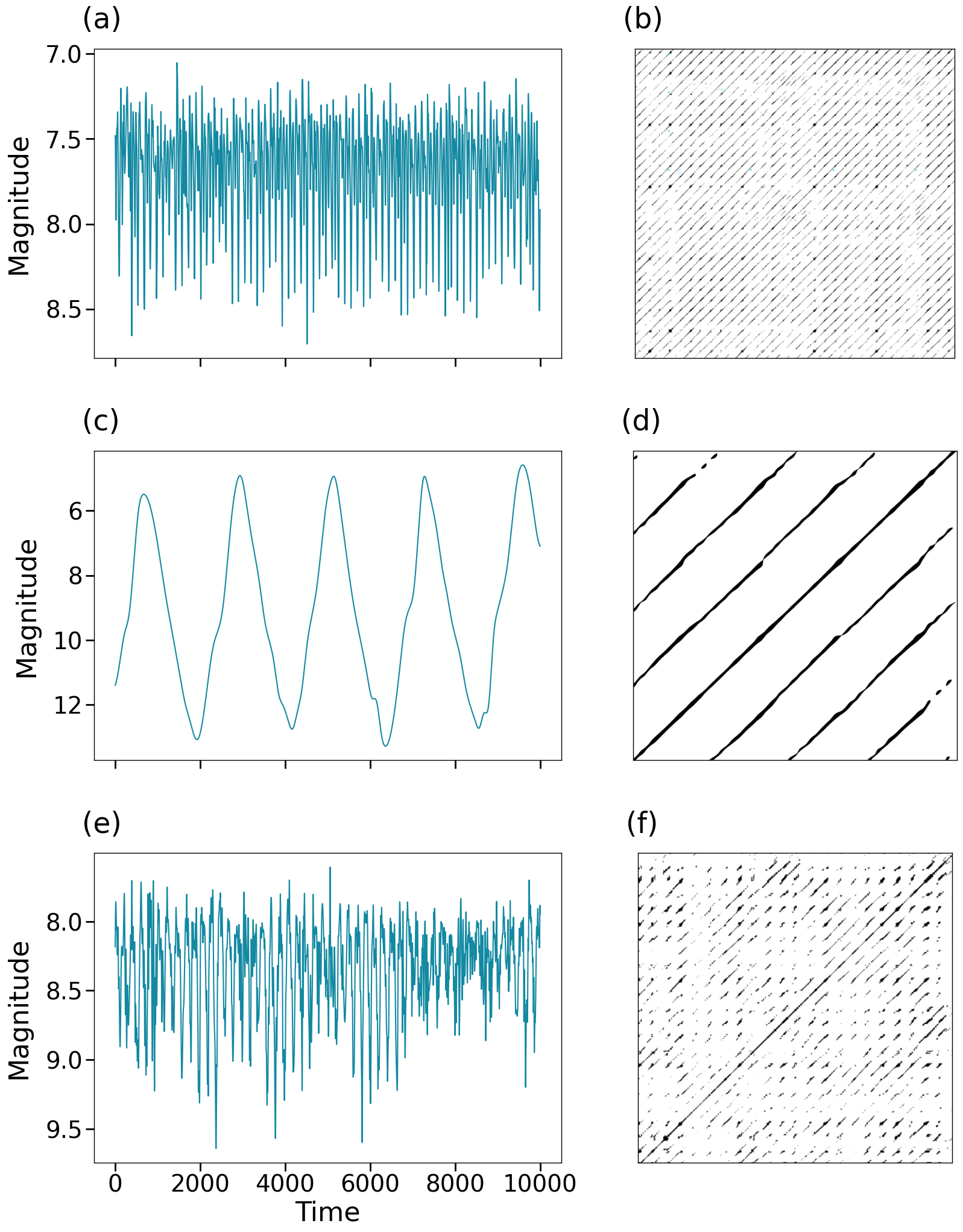}
    \caption{
       \textbf{Light curves and corresponding RPs for variable stars.}
The left column (a, c, e) shows the light curves, while the right column (b, d, f) presents the corresponding RPs for the stars AC Her, Chi Cygni, and SX Her, respectively.
    }
    \label{fig:star_lightcurve}
\end{figure}

We further validate the models using the light curve data from three variable stars, AC Her, SX Her, and Chi Cygni, sourced from the AAVSO database (\url{https://www.aavso.org/data-download}). Fig.~\ref{fig:star_lightcurve} presents the light curves and their corresponding RPs. The light curves for AC Her and SX Her span 14 years, beginning on 20 September 1980, while the Chi Cygni light curve covers a 5-year period starting on 20 September 2019. To ensure consistent sampling and reduce observational noise, we smooth the AC Her and SX Her light curves using a 5-day rolling average, and the Chi Cygni light curve using a 20-day rolling average. All light curves are then interpolated using cubic splines with a smoothing parameter of $s=0.065$. Both the proposed DBResNet-50 model and ResNet-50 correctly classify AC Her and SX Her as chaotic, consistent with earlier studies ~\cite{buchler2004chaosstars, Kollath:1997pya}. In contrast, MobileNetV2 predicts both stars to be hyperchaotic (cf. Table~\ref{tab:rr5_model_comparison_accuracy}). This suggests that lighter models may struggle to distinguish subtle differences in complex dynamical behavior compared with deeper architectures.

Finally, we evaluate the models using the reanalysis data of temperature from the NCEP (National Centers for Environmental Prediction, https://psl.noaa.gov/), focusing on two locations: Ladakh and Ranchi. The original reanalysis data, sampled four times per day, are preprocessed by binning to two samples per day, followed by removal of the dominant annual cycle. A moving mean with a window size of thirty is applied for smoothing, and the resulting signal is rescaled to the range $[0,1]$.

The DBResNet-50 model classifies both Ranchi and Ladakh as hyperchaotic. However, for Ladakh, the predicted class probabilities exhibit a mixed distribution, with hyperchaos and pink noise contributing approximately $49.7\%$ and $43\%$, respectively. For Ranchi, the model assigns about $90\%$ probability to the hyperchaos class, with all other stochastic categories receiving negligible probabilities. These results are consistent with previously reported findings, which show that for Ladakh, the RQA measures decline sharply before 1980, indicating increasingly stochastic behavior from 1980 to 2000 ~\cite{john2024recurrence}. In contrast, these measures for Ranchi remain consistently higher over the same period~\cite{john2024recurrence}.

Thus, DBResNet-50 is shown to demonstrate superior performance in classifying empirical time series data, as compared with the ResNet-50 and MobileNetV2 models. Interestingly, our model is also capable of capturing the mixed nature of the time series, as reflected in its nuanced class-probability distributions.

\begin{table*}
\centering
\caption{Performance of the deep learning models in classifying observational and experimental datasets.}
\small
\begin{tabular}{|l|c|c|c|c|}
\hline
\textbf{Time Series} &
\textbf{True Dynamics} &
\textbf{DBResNet-50} &
\textbf{ResNet-50} &
\textbf{MobileNetV2} \\
\hline

Chua Circuit 10000 & Chaotic~\cite{lakshmanan1996chaos} & Chaotic & Periodic & Chaotic \\ \hline
Chua Circuit 5000 & Chaotic~\cite{lakshmanan1996chaos} & Chaotic & Hyperchaotic & Red Noise \\ \hline
Chua Circuit 2500 & Chaotic~\cite{lakshmanan1996chaos} & Chaotic & Chaotic & Chaotic \\ \hline
GRS 1915+105 $\chi$ & White Noise~\cite{belloni2000model} & White Noise & White Noise & White Noise \\ \hline
GRS 1915+105 $\gamma$ & White Noise~\cite{belloni2000model} & White Noise & Red Noise & Red Noise \\ \hline
AC Her & Chaotic~\cite{Kollath:1997pya} & Chaotic & Chaotic & Hyperchaotic \\ \hline
Chi Cyg & Periodic~\cite{sterken1999period} & Periodic & Periodic & Periodic \\ \hline
SX Her & Chaotic~\cite{buchler2004chaosstars}  & Chaotic & Chaotic & Hyperchaotic \\ \hline
\end{tabular}
\label{tab:rr5_model_comparison_accuracy}
\end{table*}

\section{\label{sec:Discussion}Discussion}

The input representations used in this study are constructed using established concepts from nonlinear dynamical systems theory, namely phase-space reconstruction and recurrence analysis, while the deep neural network performs automated feature extraction and classification on these physically meaningful representations. As such, the proposed framework may be viewed as a theory-guided learning approach rather than a purely data-driven one. Since our objective is to capture the geometric structure of the underlying attractor, we employ time-delay embedding, which provides the standard reconstruction of multidimensional dynamics from a scalar observable. A comparison between embedded and non-embedded recurrence plots further shows that embedded representations consistently achieve superior classification performance under identical training and testing protocols. This demonstrates that phase-space reconstruction provides additional discriminative dynamical information beyond scalar recurrence representations while allowing explicit control of recurrence thresholds, which can improve robustness to noise and enhance interpretability.

The choice of recurrence plots is motivated by their direct alignment with the objective of this study, developing and validating a deep-learning framework for nonlinear time series classification based on recurrence plots images. Since recurrence plots explicitly represent pairwise recurrences in reconstructed phase space, they naturally capture the underlying geometric structure of the dynamics. Alternative representations, including phase portraits, Poincar'e sections, spectral images, Gramian Angular Fields (GAF) \cite{wang2015imaging}, and Markov Transition Fields (MTF) \cite{wang2015imaging}, encode complementary characteristics of time-series data. Rather than comparing these representations, the present work focuses on recurrence plots. The proposed architecture has therefore been evaluated only on recurrence-plot representations. Whether our proposed framework can be effectively extended to other image-based time-series representations remains an open question that requires dedicated investigation in the future.

Recurrence plots are well suited for the analysis of short and nonstationary time series because trends, drifts, and transitions are naturally reflected in their recurrence structures. To assess the robustness of the proposed framework, we introduce controlled nonstationarity into the test signals through a gradual time-varying mean drift and a time-varying amplitude (variance) modulation while keeping the trained model unchanged. The results indicate that the proposed framework maintains reasonable robustness for stochastic data under the nonstationary conditions considered, whereas the classification performance of deterministic signals gradually degrades as the strength of nonstationarity increases. Although established preprocessing techniques, such as interpolation and detrending, can mitigate the effects of data gaps and measurement noise, additional validation under conditions including missing observations, irregular sampling, and ambiguous dynamical regimes is necessary before establishing the broader applicability of the proposed framework to diverse real-world data.

We consider white, pink, and red noise because they are commonly encountered in natural and physical systems. By further evaluating unseen colored noise with power spectral density (PSD) exponents ranging from $-2$ to $0$, we find that, within the trained range of PSD exponents (0 to 2) , the predicted class probabilities vary smoothly with the spectral exponent. For such spectral exponents outside the training range, classification into specific stochastic subclasses becomes less reliable, although the signals are still predominantly classified as stochastic. Nevertheless, the present study does not establish universal generalization to all stochastic processes, such as heavy-tailed, intermittent, or strongly non-Gaussian noise, which remain important directions for future investigation.

\section{\label{sec:conclusion}Conclusion}
In this study, we show that RP images enable classification of seven different dynamical states from data without relying on computed measures using RQA. For this, we introduce a dual branch deep learning model named, DBResNet-50. The architecture of our model builds on a ResNet-50 backbone and incorporates two parallel branches, one that captures localized spatial information and another that extracts and emphasizes directional features from early mid-level feature maps, enhancing the model’s classification performance over the standard ResNet-50. The main question we address is one of broad relevance: \textit{Can we reliably infer the underlying dynamics of real-world time series data}? To achieve this, we generate RPs, extracted via phase-space reconstruction using Takens’ embedding theorem, from nonlinear time series data obtained through simulations of various dynamical systems and noise data. Our evaluated models take these RPs as input and learns to identify complex dynamical structures directly from the images of RPs.  The model thus classifies these time series into seven distinct classes: periodic, quasi-periodic, chaotic, hyperchaotic, white noise, pink noise, and red noise.

The trained DBResNet-50 model also accurately classifies unseen time series data obtained by simulating dynamical systems for example,  Duffing, and Rössler 3D systems, as well as experimental data taken from a Chua circuit, astronomical light curves from the black-hole GRS 1915+105, AC Her, SX Her, and Chi Cygni, as shown in Table~\ref{tab:dynamical_systems_extended} and Table~\ref{tab:rr5_model_comparison_accuracy}. The models also provide insight into the amount of stochasticity in the data through the class wise probabilities, helping us to interpret mixed or ambiguous regimes in the climate data as demonstrated in the case of temperature time series from Ladakh and Ranchi. We also explore the sensitivity of the empirical results by evaluating the model, trained using a 5\% recurrence rate (RR) on the same empirical data while generating RPs with 3\% and 7\% RR. We do not observe any considerable differences in the results, implying that small variations in the recurrence rate can also be accommodated by the model.

Our framework demonstrates robust classification performance across synthetic, experimental, and observational datasets. This demonstrates the model’s applicability to real-world data and reinforces the potential of using recurrence-based deep learning models for determining the inherent dynamics of the system. Compared with traditional approaches that rely on recurrence quantification measures and recurrence network metrics \cite{thakur2024mlrecurrence}, our method offers a more efficient alternative. It eliminates the need for computing recurrence measures and feature selection by learning discriminative features directly from RP images. As such, the model advances one more step in recurrence-based analysis of data.

We also compare the performance of the proposed DBResNet-50 implementation with other deep learning models, namely ResNet-50 and MobileNetV2. We find that the proposed DBResNet-50 model performs best, followed by ResNet-50, while MobileNetV2 shows comparatively lower accuracy. The RPs contain structures at multiple spatial scales, including diagonal lines, disrupted line segments, and block-like recurrence patterns associated with different dynamical regimes. The improved performance of the DBResNet-50 and ResNet-50 architectures suggests that it is effective at learning these features compared to simpler CNN models.

Looking ahead, several directions can further extend the capabilities of our model. One potential approach is to explore an ensemble Recurrence Rate (RR) method in which the model processes RPs generated with multiple recurrence rates to make a single prediction, helping bridge the gap across studies that use different RR approaches. Another interesting application involves detecting dynamical transitions within time series by segmenting RPs along the line of identity (LOI) using our model on transition-rich data to identify these substructures, similar to time-dependent RQA techniques \cite{marwan2011pitfalls}. Another enhancement includes combining our approach with model-based generative frameworks, such as delay embedding and deep delay autoencoders \cite{bakarji2023deepdelay}, to synthesize training data from partial observations of unknown systems. This could reduce dependence on fully simulated datasets and improve performance under limited data conditions.
\FloatBarrier
In conclusion, the reported study establishes and explores deep learning approaches that classify seven distinct dynamical and noise regimes directly from RPs, providing a powerful, efficient, and generalizable framework for identifying dynamical states in synthetic, astronomical, climate, and electronic circuit time series data.

\section{\label{sec:ack}Acknowledgments}
We acknowledge with thanks the variable star observations from the AAVSO International Database contributed by observers worldwide and used in this research. We also thank Prof. K. Murali for providing
Chua circuit used in validating our model. AM and CM acknowledge financial support from ANRF, India (Grant No. EEQ/2023/001080), and DST, India (INSPIRE Grant No. IFA19-PH248). AM is also thankful to Corina Simon for facilitating access to the NVIDIA A100 GPU via Google Colab.

\section{\label{sec:data}Data Availability}

The time series datasets of standard systems used in this study are generated by simulating the dynamical models listed in Tables~\ref{tab:dynamical_systems} and \ref{tab:dynamical_systems_extended} using the RK45 integrator. The experimental time series data for the Chua circuit are provided by Prof. K. Murali’s lab at Anna University, India. Time series of variable stars (light curves) are obtained from the AAVSO database (\url{https://www.aavso.org/data-download}). For the GRS 1915+105 black hole data, this research has made use of data provided by the High Energy Astrophysics Science Archive Research Center (HEASARC), which is a service of the Astrophysics Science Division at NASA/GSFC (\url{https://heasarc.gsfc.nasa.gov}). The climate data used in the study are the reanalysis data sets from NCEP (National Centers for Environmental Prediction) downloaded from \url{https://psl.noaa.gov/}.

\bibliographystyle{apsrev4-2}
\bibliography{Reference}

\end{document}